\documentclass[11pt]{article}
\usepackage[english]{babel}
\usepackage[utf8]{inputenc}
\usepackage{amsmath,amsthm,amssymb}
\usepackage{graphicx}
\usepackage{epsfig}
\usepackage{epstopdf}
\usepackage{array}
\usepackage[numbers,sort&compress]{natbib}
\usepackage{float}
\usepackage{setspace}
\usepackage{pdfpages} 
\usepackage{listings}
\usepackage{subcaption}

\begin{document}
\begin{center}
 \textbf{Probing the dynamical system and thermal behaviors of the model, $\Lambda_0+3 \beta H^2.$}\\[0.1in]
 Jayadevan A P,$^1$ Mukesh M$^{2}$, Shaima Akbar$^3$,  and Titus K Mathew$^4$\\
 $^{1,2,3}$Department of Physics, Sree Kerala Varma College, Thrissur, India \\
 $^4$Department of Physics, Cochin University of Science and Technology, Kochi-22,India. \\
 apjaps@gmail.com,mukeshmurali1994@gmail.com,shaimaakbar333@gmail.com  titus@cusat.ac.in
\end{center}

\abstract{In this work we study the cosmological evolution of a two component model with non-relativistic dark matter and decaying vacuum of the 
form $\Lambda = \Lambda_0 + 3 \beta H^2.$ We contrast the model the model with the supernovae data and found that the model parameter $\beta=-0.010$ 
when the interaction parameter $b=0$ and is $\beta=-0.002$ when $b=0.001.$ The thermal evolution study of the model indicates that it obeys the 
generalized second of entropy and also satisfies the convexity condition, $\ddot S <0$ so that the model behaves like an ordinary macroscopic system which 
approaching a stable equilibrium thermal state is the asymptotic condition. The dynamical system analysis reveals that, the model posses a prior 
decelerated epoch represented by a saddle critical point and the late critical point represents an accelerating epoch which posses a convergence of 
phase-space trajectories form both the regions of the point hence can be considered as stable.}

\section{Introduction} 
The discovery of the recent acceleration in the expansion of the universe\cite{Riess,Perlmutter} become the most important 
theoretical challenge now a days. Dark energy become the most promising candidate for understanding this, but
 the nature of this exotic cosmic component\cite{sami,sahni} is still a mystery.
The standard $\Lambda$CDM model, in which the cosmological constant is the dark energy, become the most 
successful model in explaining this recent acceleration. 
However this model faces with certain serious issues such as cosmological constant problem and coincidence problem\cite{sami}. 
The cosmological constant problem is regarding the 
large discrepancy between the observed value of cosmological constant and its theoretical prediction from 
quantum field theory\cite{Weinberg1}, which is nearly 
$120$ orders of magnitude. Coincidence problem is about the mysterious coincidence of the present densities of dark matter 
and dark energy.  
The approach to solve these problems is to consider that the dark energy density is varying with the  
expansion of the universe. Since then various dynamical dark energy models have been proposed, for instance, scalar field model\cite{sami}, 
the holographic Ricci dark energy model\cite{Cao1,Feng1,Prasee1,maia}, which is arises from the cosmological holographic 
principle\cite{Suss1} 
and Chaplygin gas models\cite{Kamen1} are a few. 

Recently decaying vacuum \cite{carvalho,lima}, also called as the running vacuum, has been considered as a suitable  
dynamical dark energy candidate. It has plausibility of giving a unified description of the  
early inflation and the late acceleration\cite{Joan1}.
Some recent analysis on cosmological data, gives strong indications for such a slowly varying vacuum energy component\cite{Valet1}.
The decaying vacuum models has its root in quantum field theories\cite{Sola1,Sola2}. For example 
it can be realized through scalar fields\cite{Wetterich1,Ratra1,Caldwell1} coupling  with the dark matter component. 
There are models of decaying vacuum in recent literature which dealing with the possible particle production arised due to 
the coupling between the dark sectors\cite{Alcaniz1}. 

Decaying vacuum 
 based on the renormalization group  approach in quantum field theory in curved space-times has been considered in 
 references\cite{Nelson1,Buch1} and it turn 
 out to be a suitable candidate. 
The effective action in such theories inherits quantum effects from the matter sector. In general, the renormalization group 
techniques in curved space-times leads to a vacuum energy depending on the Hubble parameter $H$ and its time derivative, of the form 
$\rho_{\Lambda}(H, \dot H) = M_{Pl}^2 \Lambda(H, \dot H)$ (see \cite{Sola3,Sola4,Shapiro1} and references there in).
Motivating from this many have considered varied phenomenological versions of decaying vacuum
\cite{Bertolami1,Ozer1,Peebles1,carvalho,lima,Overduin1}. One such simple but potential candidate for the decaying vacuum is
 $\Lambda(t)=\Lambda_0+3\beta H^2$\cite{Bessada1}, where
$\Lambda_0$ is a bare cosmological constant and $\beta$ is the model parameter.
The $\Lambda_0$ term is crucial in predicting the transition from a prior decelerated epoch to a later 
accelerated epoch\cite{Basilakos1,Paxy1}, in the absence of which the model is either eternally accelerating or decelerating. 
This model 
has been used to study 
the star formation rate and have found a constraint to the parameter as,
$\beta<0.01$\cite{Bessada1}. The real significance of this model of decaying vacuum is in the context of recent 
acceleration, in which it deserve a detailed study. Also using the star formation data to constrain parameter $\beta$ is actually in tension, since 
it has been pointed out that star formation data has large 
observational uncertainties, especially for high redshift data. On the other hand, there is comparatively less uncertainty in the 
supernovae data\cite{Kistler1}. So it is worth 
analyzing the model in the light of the supernovae data.
In the present work we have 
studied the cosmological evolution of this model of decaying vacuum with reference to the supernovae data.
We have extended our analysis by considering the possible interaction between decaying vacuum and the 
dark matter. We also perform a study on the thermodynamics and dynamical system behavior of the model.

The paper is organized as follows. 

\section{Background evolution of the model}

We consider a flat FLRW universe, with cold dark matter and decaying vacuum of the form\cite{Bessada1} 
\begin{equation}\label{eqn:F1}
 \Lambda(t) = \Lambda_0 + 3 \beta H^2
\end{equation}
as cosmic components. The evolution of the FLRW universe satisfies the basic Friedmann equation,
\begin{equation}\label{eqn:Fried1}
 3 H^2 = 8\pi G \left(\rho_m+\rho_{\Lambda} \right),
\end{equation}
where $\rho_m$ is the density of the non-relativistic matter, mainly the cold dark matter and 
$\rho_{\Lambda}=\Lambda(t)/8 \pi G$ 
is the decaying vacuum energy density.
The cosmic 
components together satisfies the energy and momentum conservation,
 $T^{\mu\nu}_{;\nu}=0,$
where $T^{\mu\nu}=\left(\rho+_P \right)u^{\mu}u^{\nu} - p g^{\mu\nu},$ with $u^{\mu}$ is the fluid 
four velocity, 
$\rho$ and $P$ are effective fluid density and pressure respectively. In the absence of source and sink, this
 implies a conservation law of the form,
\begin{equation}\label{eqn:con1}
 \dot{\rho}_m+\dot{\rho}_\Lambda+3H\left(\rho_m+p_m\right)+3H\left(\rho_{\Lambda}+p_{\Lambda}\right) =0
\end{equation}
where $p_m$ and $p_{\Lambda}$ are the pressures due to non-relativistic matter and decaying vacuum 
respectively and the over dot represents a derivative with respect to cosmic time. For  
non-relativistic matter, $p_m \sim 0$ and for decaying vacuum, $p_{\Lambda}=-\rho_{\Lambda}.$ As a result the above conservation equation become,
\begin{equation}\label{eqn:con2}
\dot{\rho}_m+\dot{\rho}_{\Lambda}+3H\rho_m=0
\end{equation}
From equations (\ref{eqn:F1}) and (\ref{eqn:con2}) together with a change of variable from $t$ to 
$x=\log a,$ where $a$ is the scale factor of expansion, we have,
\begin{equation}
\frac{dh^2}{dx}+3(1-\beta)h^2+3\Omega_{\Lambda_0}=0
\end{equation}
where $h=H/H_0, \Omega_{\Lambda0}=\frac{\Lambda_0}{3H_0^2}$ and $H_0$ is the current value of the Hubble parameter. 
Solving this, we get the Hubble parameter as,
\begin{equation}\label{eqn:Hubble2}
 H^2=H^2_0\left[\left(1-\frac{\Omega_{\Lambda 0}}{1-\beta}\right)a^{-3(1-\beta)}+\frac{\Omega_{\Lambda 0}}{1-\beta}\right]
\end{equation}
For $\beta=0$ the above equation reduces to the corresponding equation in $\Lambda$CDM model as expected.
In the limit $a\to 0$, the Hubble parameter, 
$H \to (1-\frac{\Omega_{\Lambda 0}}{1-\beta})a^{-3(1-\beta)}$ and it implies a prior decelerated matter dominated epoch.
While in the future asymptotic limit, $a\to \infty$, the Hubble parameter attain a constant value, 
$H \to H_0\sqrt{\frac{\Omega_{\Lambda 0}}{1-\beta}}$ and 
it corresponds to the end de Sitter epoch at which the expansion is accelerating. So the model does implies a transition from a
decelerated phase to a late accelerated phase of expansion. It is evident that, for $\Lambda_0=0$ the model does not predict a 
transition from a prior decelerated to a late accelerated phase. The evolution of the Hubble parameter for the best fit 
values of the model parameters, 
$\Omega_{\Lambda0}=0.685, \, H_0=69$ and $\beta=-0.010$ (estimated in a later section) is given in Fig.\ref{fig:Hplot1}, 
in comparison with the that of standard $\Lambda$CDM model. The evolution $H$ is almost same in both models.

The scale factor of expansion can then be obtained by integrating the Hubble parameter and the result is,  
\begin{equation}
\begin{split}
a={\left(\frac{1-\beta-\Omega_{\Lambda 0}}{\Omega_{\Lambda 0}}\right)}^{\frac{1}{3(1-\beta)}} \hspace{1in} \\ 
\left(\sinh\left[\frac{3}{2}\sqrt{(1-\beta)\Omega_{\Lambda 0}}H_0 t\right] \right)^{\frac{2}{3(1-\beta)}}
\end{split}
\end{equation}
where $t$ is the cosmic time. First of all, this equation reduces to the corresponding equation of $\Lambda$CDM model for 
$\beta=0.$ 
For small values of $t,$ the scale factor evolves as, $a \propto t^{\frac{2}{3(1-\beta)}}$ and is corresponding to the prior 
matter dominated phase of the universe, while for large values of $t$, the scale factor,
$a \propto \exp\sqrt{\frac{\Lambda_0}{3(1-\beta)}}t,$ implying the later de Sitter phase. The evolution 
of the scale factor in contrast to the standard model $\Lambda$CDM is shown in Fig. \ref{fig:Hplot1}. As like the Hubble parameter 
the scale factor evolution of the model is also very close to that of the $\Lambda$CDM model.

\subsection{Model with explicit interaction between the dark sectors}

It could be possible to have non-gravitational interaction between the dark sectors of the universe. The
conservation equations with interaction will be\cite{Prasee1}
\begin{equation}\label{eqn:con22}
\dot{\rho_m}+3H(\rho_m+p_m)=Q, \qquad 
\dot{\rho_\Lambda}+3H(\rho_\Lambda+p_\Lambda)=-Q
\end{equation}
 The term $Q$ represents the interaction between the dark sectors and if $Q>0$ there occur energy transfer from dark energy to the dark 
 matter sector and if $Q<0$ energy transition takes place from matter sector to dark energy. Owing to the lack of 
detailed knowledge of this interaction we will assume a phenomenological form for $Q$ as\cite{Prasee1},
\begin{equation}
Q=3Hb\rho_m
\end{equation}
where $b$ is a parameter characterizing the interaction, the value of which is to be constrained by the observation data.

The pressure due to the non-relativistic matter is zero hence the first of the conservation equations (\ref{eqn:con22}) become
\begin{equation}
\dot{\rho}_m = - 3(1-b)H\rho_m 
\end{equation} 
Using this equation and the Friedmann equation (\ref{eqn:Fried1}), it can be easily shown that,
\begin{equation}
 \dot H = -\frac{\gamma}{2} \rho_m
\end{equation}
where $\gamma = \frac{1-b}{1-\beta}.$ Substitute for $\rho_m$ and subsequently for $\rho_{\Lambda},$ we get
\begin{equation}
\frac{d(H/H_0)^2}{dx}=-3\gamma (1-\beta) (H/H_0)^2 + 3 \gamma \Omega_{\Lambda 0}
\end{equation} 
Here we have changed the variable form cosmic time $t$ to $x=\log a.$
On solving this we get the Hubble parameter as,
\begin{equation}\label{eqn:d}
H^{2}=H_0^{2} \left[ \left(1-\frac{\Omega_{\Lambda0}}{1-\beta} \right) a^{-3(1-b)} + \frac{\Omega_{\Lambda0}}{1-\beta} \right]
\end{equation}
It should be noted that, for $b=0$ the above equation would not reduces to the Hubble parameter in 
equation (\ref{eqn:Hubble2}). Because in such case, the conservation equations (\ref{eqn:con22}) becomes the one corresponding 
to the independent conservation of dark matter and dark energy respectively. Dark energy represented by the running vacuum 
in equation (\ref{eqn:F1}) is self conserved only for $\beta=0,$ for which the the running vacuum become equal to the 
bare constant, $\Lambda_0.$ Hence $b=0$ naturally implies a vanishing $\beta,$ correspondingly the Hubble parameter 
reduces the pure $\Lambda$CDM model, $H^2/H_0^2=(1-\Omega_{\Lambda0})a^{-3}+\Omega_{\Lambda0}=\Omega_{m0}a^{-3}+\Omega_{\Lambda0}.$ 

The asymptotic behavior of the Hubble parameter (\ref{eqn:d}) is as follows. In the limit $a \to 0$ the Hubble behave as 
  $H^2/H_0^2 \to (1-\frac{\Omega_{m0}}{1-\beta})a^{-3(1-b)},$   
 representing the matter dominated decelerating epoch. But as $a \to \infty$ the Hubble parameter go as, 
 $H^2/H_0^2 \to \frac{\Omega_{\Lambda 0}}{1-\beta} ,$ corresponds to the late accelerating epoch dominated by dark energy. Thus the model gives a transition from an early decelerated phase to a later accelerated phase as warranted by the observation.
The analytical expressions for scale factors is obtained by integrating the Hubble parameter as,
\begin{equation}
\begin{split}
a={\left(\frac{\Omega_{\Lambda_0}}{1-\beta-\Omega_{\Lambda_0}}\right)}^{\frac{-1}{3(1-b)}} \hspace{1in} \\
\sinh^{\frac{2}{3(1-b)}}\left[\frac{3}{2}\sqrt{\frac{\Omega_{\Lambda 0}}{1-\beta)}}(1-b) H_0t\right].
\end{split}
\end{equation}
For small values of $t,$ the scale factor varies as $a \propto t^{\frac{2}{3(1-b)}}$ and for large values of $t,$ 
scale factor shows exponential variation of time as $a \propto \exp\sqrt{\frac{\Lambda_0}{3(1-\beta)}}t$ . 
Earlier case implies the matter dominated epoch while later case indicate the de Sitter phase of the universe.

The evolution of the Hubble parameter in terms of cosmic time is obtained form equation(\ref{eqn:d}) as,
\begin{equation}\label{eqn:Hwithb1}
H=H_0\sqrt{\frac{\Omega_{\Lambda_0}}{1-\beta}}\coth\left[\frac{3}{2}H_0t(1-b)\sqrt{\frac{\Omega_{\Lambda_0}}{(1-\beta}}\right]
\end{equation}

\section{Estimation of model Parameters and evolution of the cosmological parameters}
In this section we are evaluating the model parameters, especially, $\beta$ using the supernova data. As we have mentioned in 
the introduction, the parameter $\beta$ has been evaluated before in \cite{Bessada1} for $b=0$  
 with reference to the cosmic star formation data.
The cosmic star formation data are exist for relatively high redshift. For instance there exists cosmic 
star formation data for large redshift of the oder of $z \sim 10,$ which has been obtained form high redshift galaxies and some 
gamma ray bursts\cite{Kistler1}. However the observational uncertainties are considerably large for redshifts, $z>3.$ So the authors 
in reference\cite{Bessada1} were able to find only an upper bound to the parameter $\beta.$ The supernovae data has got comparatively 
less uncertainties primarily due to very fact that the corresponding data were obtained for redshifts less 2 and also they are available in sufficiently 
large numbers. More over Supernovae 
data were give sufficient information regarding the era of accelerated expansion of the universe and hence most suitable than cosmic 
star formation formation data to constrain the dark energy parameters.
 
The parameters are extracted by the method of $\chi^{2}$ analysis. Luminosity distance $d_L$ of a supernova at a redshift $z$ is given by
\begin{equation}
d_L=c(1+z)\int_0 ^{z} \frac{dz^{\prime}}{H}
\end{equation}
where $c$ is the velocity of light.
From this the distance modulus $\mu_t$ of the star can be calculated using the relation,
\begin{equation}
\mu_t=5log_{10}\left[\frac{d_L}{Mpc}\right]+25
\end{equation}
This theoretical estimate of the modulus can then be compared with the corresponding observational value $\mu_i$ for 
the same redshift. Such comparison for the the observational supernovae data are weighted with the respective standard error, $\sigma_i$ in the observation.
Then the total $\chi^{2}$ function which compare these two moduli can be defined as,
\begin{equation}
\chi^{2}=\sum^{n}_{i=1}\frac{(\mu_t-\mu_i)^2}{\sigma_i^{2}}
\end{equation}
where $n$ is the total number of supernovae observational data.
\begin{table}[h]
\centering 
\caption{Best fit values of the parameters of the model using Supernovae data}
\begin{tabular}{llllll}
 \hline\noalign{\smallskip}
 $b$ & $\chi^2_{min}$ & $\chi^2_{min}/d.o.f$ & $H_0$ & $\Omega_{\Lambda 0}$ & $\beta$ \\
 \noalign{\smallskip}\hline
  - & 315.337 & 1.03 & 69 &  0.685 & -0.010 \\
  0.001 & 312.478 & 1.02 & 70 & 0.713 & -0.002 \\
 \noalign{\smallskip}\hline 
 \label{tab:t1}
 \end{tabular}
\end{table}
We have used the Union data consists of 307 supernovae observations\cite{Kowa1} in the redshift range $0.01<z<1.5.$ We have 
evaluated the parameters for model with no 
\begin{figure}
  \centering
  \includegraphics[scale=.5]{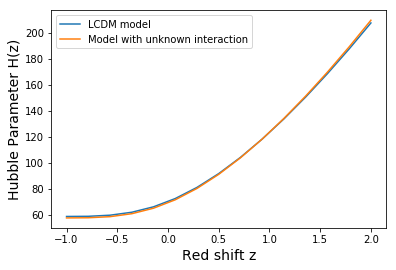}, \qquad 
  \includegraphics[scale=.5]{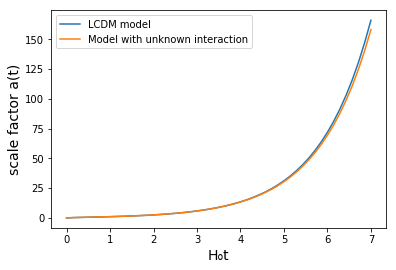}
  \caption{\footnotesize{ Comparison of evolution of Hubble parameter(left panel) and scale factor (right panel) 
 between $\Lambda$CDM and the present model. Both are having almost the same evolution.}}
  \label{fig:Hplot1}
\end{figure}
\begin{figure}
\centering
\includegraphics[scale=0.5]{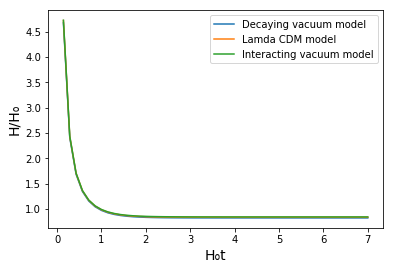} \qquad 
 \includegraphics[scale=.5]{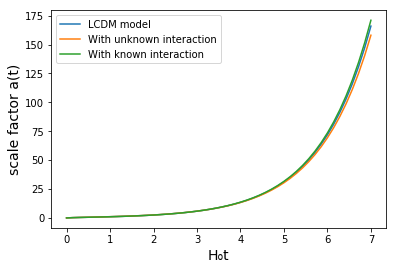}
\caption{Evolution of the Hubble parameter and corresponding scale factor with interaction parameter $b=0.001.$}
\label{fig:Hb1}
\end{figure}
known interaction and also model with known interaction parameter $b$ and the results are shown in Table.\ref{tab:t1}.
In contrast to the reference \cite{Bessada1}, we have 
evaluated the exact value of the parameter $\beta$ and and found to be negative, however it is compatible with the upper limit 
evaluated in \cite{Bessada1} using the star formation data.  For non-zero value of the interaction 
parameter $b=0.001,$ the cosmological parameters have slightly higher value, especially the value $\beta=-0.002$ at which the present value 
of the Hubble parameter is slightly less.

The evolution of the Hubble with redshift for the case where the cosmic components follow the common conservation law, 
is given in the left panel of Fig.\ref{fig:Hplot1} corresponding to the best fit 
values of the parameters. For comparing with the $\Lambda$CDM model we have plotted the corresponding $H$ parameter in the same 
panel and is almost coincide that of the present model. This indicates that the model 
is similar to $\Lambda$CDM as far as the expansion is concerned. In the right panel of the same figure we have given the evolution 
of the scale factor with $H_0t.$ Side by side with it the corresponding plot for the standard $\Lambda$CDM is also 
given for comparison. First the plot indicating the presence of prior deceleration and a later acceleration in the expansion and 
secondly it says that the model under consideration is giving a behavior almost like that of the $\Lambda$CDM model. The similarity 
in the behavior of the present decaying model is evidently due to the low value of the model parameter $\beta.$

 The time 
 evolution of the Hubble parameter for non-zero interaction parameter for the best estimated values of the parameters 
 is plotted in left panel of Figure.\ref{fig:Hb1} and 
  evolution of the scale factor is given in the right panel.
There is no substantial difference in the behavior of the Hubble parameter as compared to $\Lambda$CDM model.
In the asymptotic limit when the universe approach a de Sitter epoch, the cosmological parameter will tends to 
a constant value, $\frac{\Omega_{\Lambda0}}{1-\beta}.$ The best estimate of the parameter, $\beta=-0.002,$ 
the values become, $\Omega_{\Lambda}=\frac{\Omega_{\Lambda0}}{1.002}$  and  is very close to $\Omega_{\Lambda0}.$ 
Most important factor to be noted is that, unlike in the $\Lambda$CDM model, even though the vacuum energy corresponds 
evolving,  the evolution of the other cosmological parameters like Hubble parameter and scale 
factor are in concordance with that in the standard $\Lambda$CDM model.

\section{GSL and entropy boundedness
}

In this section we analyses the thermodynamic evolution of the model.
Considering the Universe as an isolated macroscopic system, the total entropy comprises the entropies due to  dark energy, 
dark matter plus that of horizon. The entropy evolution of the fluids within the horizon of the universe is 
is given by the Gibbs relation,
\begin{equation}\label{eqn:Tds1}
TdS=pdV+dU,
\end{equation}
where $T$ is the temperature, $V$ is the volume of the universe enclosed by the horizon and $U=\rho V$ is the 
internal energy of the cosmic fluids. Along with this the entropy also satisfy,\cite{Mazum1}
\begin{equation}
 \frac{\partial^2 S}{\partial T \partial V} = \frac{\partial^2 S}{\partial V \partial T}
\end{equation}
which actually follows from the integrability condition. This immediately implies,
\begin{equation}
 dS = d\left[\frac{(\rho+p)V}{T}+C \right]
\end{equation}
where $C$ is the integration constant. The term, $(\rho+p)/T$ can be interpreted as the entropy density. 
These equations directly implies that, the 
\begin{figure}[ht]
 \centering
\includegraphics[scale=0.5]{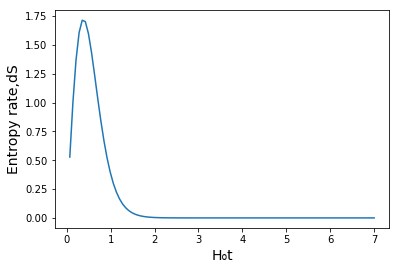}
 \caption{\footnotesize{Variation of  rate of change of total entropy $\dot{S}$ with $H_0t.$}}
 \label{fig:sevol1}
 \end{figure}
entropy contribution from the decaying vacuum, since it satisfying the equation of state $p_{\Lambda}=-\rho_{\Lambda},$ is zero. 
So only non-relativistic matter and horizon, contribute to the total entropy. 
The generalized second law (GSL) then demands that, the total entropy, due to the matter and horizon together, must always increase, that
is,
\begin{equation}\label{eqn:gsl1}
 \dot S=\dot{S}_m + \dot{S}_h \geq 0,
\end{equation}
where the over-dot represents a derivative with respect to time.
On the other hand, boundedness of the entropy, at which entropy attains a maximum, leads to the convexity condition, $\ddot S < 0$ at least at the end stage of the 
evolution. This condition also guarantees the achievement of stable equilibrium asymptotically.

\subsection{For the case where the cosmic components satisfy a common conservation law}
Here we check the status of the GSL and the convexity condition for case in which the cosmic components conserve together and the Hubble 
parameter is given by the equation(\ref{eqn:Hubble2}).
The rate of change of entropy due to non-relativistic matter can be obtained using the equation (\ref{eqn:Tds1}) as,
\begin{equation}\label{eqn:smat1}
 T\dot{S}_m=\dot{\rho}_m V+\dot{V}\rho_m
\end{equation} 
where
 $V=\frac{4\pi}{3H^3}$ is the Hubble volume, since we assume Hubble horizon as the thermodynamic boundary. 
 The horizon entropy is given the Bakenstein relation for horizon entropy, $S_h=A/4G,$ where 
 $A=4\pi/H^2$ is the area of the horizon. On taking the time derivative $S_h$ and on multiplying it with the horizon temperature,
 $T=H/2\pi,$ we get,
 \begin{equation}\label{eqn:hmat1}
  T\dot{S}_h=-8\pi \frac{\dot H}{H^2},
 \end{equation}
where we have assumed equilibrium between the cosmic components and the horizon with a uniform temperature, $T$ and 
we have adopted the standard units such that, $8\pi G=1.$ Adding
equations (\ref{eqn:smat1}) and (\ref{eqn:hmat1}) to get the total entropy rate and arrange them suitably, we get,
\begin{equation}\label{eqn:sdot12}
 \dot S =\dot{S}_m + \dot{S}_h = \frac{2\pi}{H}\left(-4\pi(3-\beta)+\frac{4\pi \Lambda_0}{H^2}\right)\frac{\dot{H}}{H^2},
\end{equation}
where we have used the Friedmann equation and the expression for the decaying vacuum. Since $\dot{H}<0,$ as the Hubble parameter 
is decreasing as the universe expands and the decaying 
vacuum has a quintessence nature as the equation of state is always remains at $-1,$ the GSL condition 
in equation (\ref{eqn:gsl1}) is 
satisfied if,
\begin{equation}
 \frac{\Lambda_0}{H^2} \leq (3-\beta),
\end{equation}
which can be suitably translated in to,
\begin{equation}
 3\Omega_{\Lambda 0} \leq (3 - \beta) h^2.
\end{equation}
For best estimated values, $\Omega_{\Lambda 0}=0.689, \, \, \beta=-0.010,$  the above condition is satisfied for the present state 
of the universe, where $h=1.$ 
For checking about the general validity of GSL we have plotted the evolution of 
$\dot S$ with cosmic time and is given in figure.\ref{fig:sevol1}, showing that
GSL is satisfied through out the evolution of the universe.

\begin{figure}
  \centering
  \includegraphics[scale=.5]{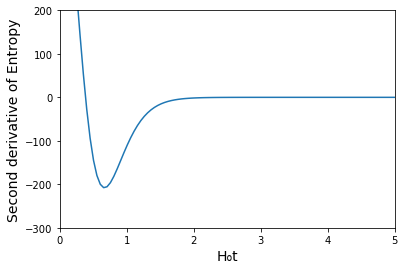}
  \caption{\footnotesize{Variation of  second derivative of total entropy $\ddot{S}$ against $H_0 t$}}
  \label{fig:dds1}
\end{figure}

The validity of GSL however does not guarantee that the entropy 
approaches a maximum at the end stage. The attainment of maximum entropy at the end stage indicates that the 
system approaches a stable equilibrium at which it satisfies 
the convexity condition, $\ddot S<0$ \cite{Diego1}.
From equation (\ref{eqn:sdot12}), we have the second derivative of entropy as,
\begin{equation}
\begin{split}
\ddot{S}=24\pi^2\Omega_{\Lambda 0} H_0^2\left[\frac{\ddot{H}}{H^5}-\frac{5\dot{H}^2}{H^6}\right]+ \\ 24\pi^2\left[\frac{\ddot{H}}{H^3}-
\frac{3\dot{H}^2}{H^4}\right]\left(\frac{\beta}{3}-1\right).
\end{split}
\end{equation}
Substituting for the second derivative of $H$ by using Friedmann equation as, $\ddot H=-3(1-\beta)H\dot H,$ leads to an equation 
in terms of $\dot H$ and $H,$
\begin{equation}
\begin{split}
\ddot{S}=24\pi^2\Omega_{\Lambda 0} H_0^2\left[\frac{3(\beta-1)\dot{H}}{H^4}-\frac{5\dot{H}^2}{H^6}\right]+ \\ 24\pi^2
\left[\frac{3(\beta-1)\dot{H}}{H^2}-\frac{3\dot{H}^2}{H^4}\right]\left(\frac{\beta}{3}-1\right).
\end{split}
\end{equation}
The behavior of second derivative of entropy with $H_0 t$ is shown figure \ref{fig:dds1}, form which 
\noindent it is clear that the second derivative of entropy approaches zero from below in the 
long run of the universe. Hence the convexity condition is satisfied in the last stages and as a result
entropy will never grow unbounded in this case. The equilibrium at the last stage is hence a thermodynamically stable one.

\subsection{For the case with interaction $Q=3Hb\rho_m$ }
We now check the status of the GSL and the convexity condition for the case where the interaction between the cosmic 
components is introduced through a coupling parameter, $b.$ For checking the status of the GSL, we will use equation(\ref{eqn:sdot12}) 
which will lead to the same condition in the present case also and it guarantee the validity of GSL.
    \begin{figure}
        \centering
        \includegraphics[scale=0.5]{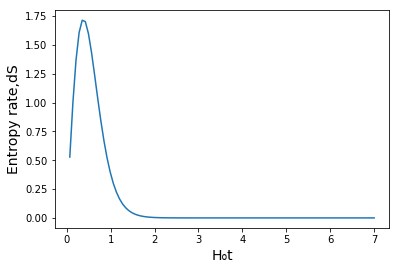}
        \caption{\footnotesize{Variation of  rate of change of total entropy $\dot{S}$ with $H_0 t$ }}
    \end{figure} 
The second derivative of entropy is then obtained as, 
\begin{equation}
\begin{split}
\ddot{S}=24\pi^2\Omega_{\Lambda 0} H_0^2\left[\frac{3(b-\gamma)\dot{H}}{H^4}-\frac{5\dot{H}^2}{H^6}\right]+ \\ 
24\pi^2\left[\frac{3(b-\gamma)\dot{H}}{H^2}-\frac{3\dot{H}^2}{H^4}\right](\frac{\beta}{3}-1),
\end{split}
\end{equation}
The figures below shows the second derivative of entropy variation with the the cosmic time $H_0 t.$ 
\begin{figure}
  \centering
  \includegraphics[scale=.5]{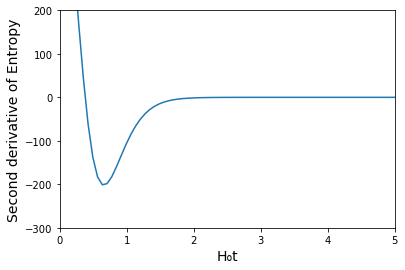}
  \caption{\footnotesize{Variation of  second derivative of total entropy $\ddot{S}$ against $H_0 t$}}
\end{figure}
Here also, as in the previous case, the second derivative of the entropy is approaching zero from below 
at the end stage, implies that the 
convexity condition is satisfied asymptotically, consequently the entropy is bounded.
This otherwise implies that the universe tending towards a state of thermodynamically stable equilibrium.

\section{Phase space analysis}
 In the previous section, the thermodynamic analysis have shown that, the model is thermodynamically stable as the entropy 
satisfies the generalized 
second law and also it satisfies the convexity condition. So it is worth studying the dynamical system behavior of the model, 
and it's asymptotic behavior.
 The method is to construct suitable autonomous differential equations from the Friedmann equations governing the time evolution of the model.
  Then the critical points corresponding to these autonomous equation can be analyzed for the stability conditions.
  
  The dynamical variables 
 selected for constructing the autonomous equations are,
 \begin{equation}
 u=\frac{\rho_m}{3H^{2}}, \hspace{0.5in} 
 y=\frac{\rho_\Lambda}{3H^{2}}.
\end{equation}
Both these varies in the range, $0\leq u\leq 1$ and $0\leq y\leq 1$.
We then have the autonomous differential equations, using Friedmann equations, as 
\begin{equation}\label{eqn:e}
u^{\prime}=3u(b-\gamma)-3(1-y)\left(\frac{b-\gamma}{1-\beta}\right)\left(1-\beta-\Omega_{\Lambda0}\frac{H_0^{2}}{H^{2}}\right)
\end{equation}
\begin{equation}\label{eqn:f}
y^{\prime}=-3(1-y)b-3y\left(\frac{b-\gamma}{1-\beta}\right)\left(1-\beta-\Omega_{\Lambda0}\frac{H_0^{2}}{H^{2}}\right),
\end{equation}
where the prime denote a derivative with respect to $'\textrm{log} \, a'$. The critical points $(u_c,y_c)$ are found  equating both
the equations to zero. Since we are interested in the asymptotic behavior we will select the critical points corresponding initial 
and final asymptotic condition and are,\\
\begin{equation}
\begin{split}
 (u_c,y_c) = (1,0) \hspace{0.3in} \textrm{corresponding to} \, H \to \infty (a \to 0) \\
 (u_c,y_c) = (1,1) \hspace{0.3in} \textrm{corresponding to} \, H \to 0 (a \to \infty)
 \end{split}
\end{equation}

To check the 
stability of the system around the resulting critical points, we consider small perturbation around the critical point and linearize the dynamic system. 
Perturbation around critical point can be written as $u\rightarrow u_c +\delta u$ and $y\rightarrow y_c +\delta y$. The linearization leads to
the following matrix equation,
\begin{equation}
	\begin{bmatrix}
	\delta u^{\prime} \\
	\delta y^{\prime}
	\end{bmatrix}
	=
	\begin{bmatrix}
	\left(\frac{\partial f}{\partial u}\right)_0 & \left(\frac{\partial f}{\partial y}\right)_0\\
	\left(\frac{\partial g}{\partial u}\right)_0 & \left(\frac{\partial g}{\partial y}\right)_0
	\end{bmatrix}
	\begin{bmatrix}
	\delta u \\
	\delta y
	\end{bmatrix}
\end{equation} 
 where the derivatives are evaluated at the critical points. From this matrix equation we can write the Jacobian matrix corresponding to the autonomous 
 equations Eq.(\ref{eqn:e}) and Eq.(\ref{eqn:f})
 \begin{equation}
 	\begin{bmatrix}
	3(b-\gamma) & 3(\frac{b-\gamma}{1-\beta})(1-\beta-\Omega_{\Lambda0}\frac{H_0^{2}}{H^{2}}) \\
	0 & 3b-3(\frac{b-\gamma}{1-\beta})(1-\beta-\Omega_{\Lambda0}\frac{H_0^{2}}{H^{2}}) 
	\end{bmatrix}
 \end{equation}
Eigen values are obtained by solving the corresponding Jacobian matrix are
\begin{equation}
\begin{split}
\lambda_1=-3, \quad \lambda_2=3 \quad \quad \textrm{cooresponding to} \, (u_c,y_c)=(1,0) \\
\lambda_1=-3, \quad  \lambda_2=0 \quad \quad \textrm{cooresponding to} \, (u_c,y_c)=(0,1)
\end{split}
\end{equation}
For $(u_c,y_c)=(1,0)$, both eigen values are real and are opposite in sign, hence the critical point is an unstable saddle point. 
Hence any trajectory starting form the neighborhood 
 \begin{figure}[ht]
 \includegraphics[scale=.40]{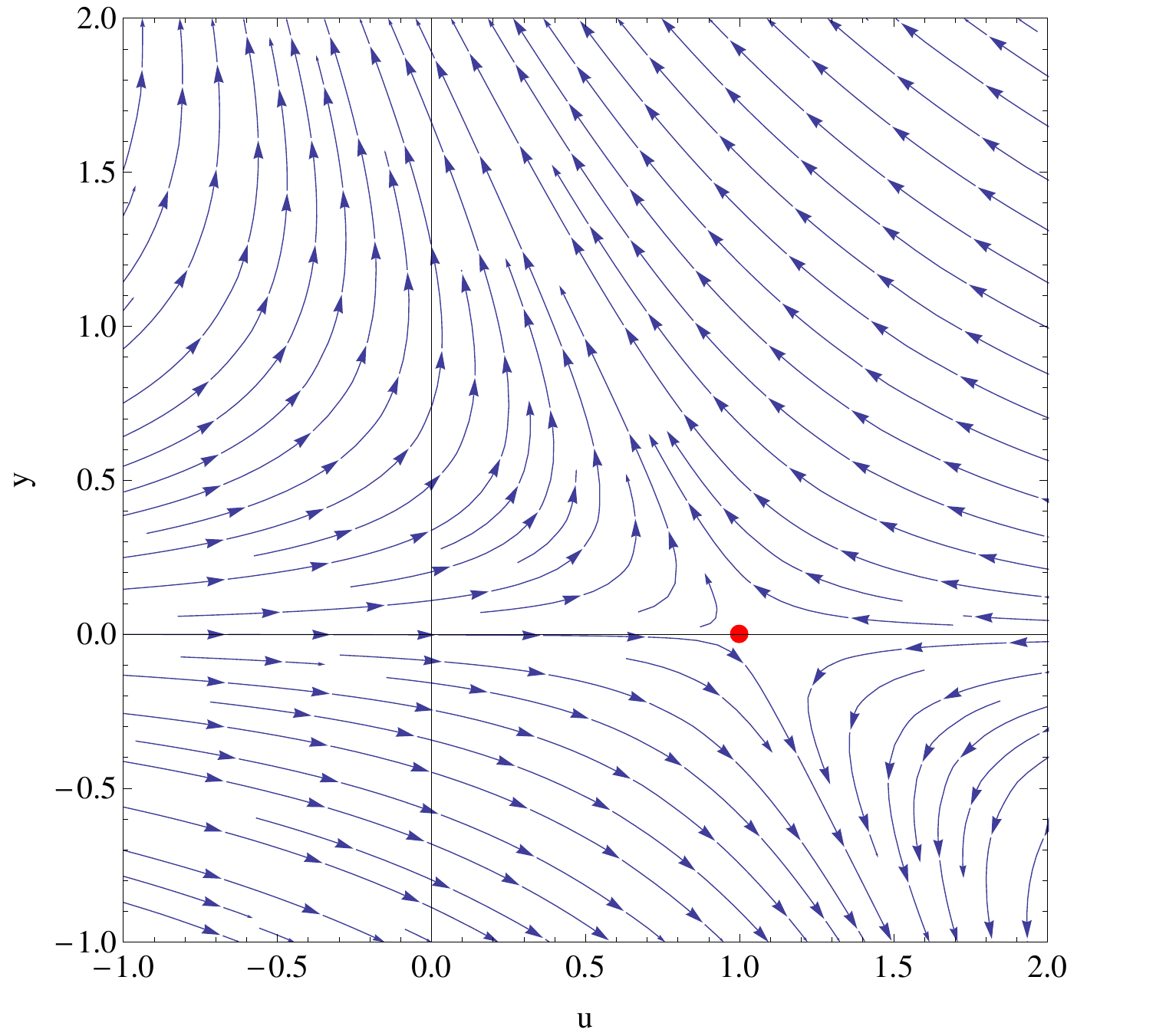}
 \includegraphics[scale=.47]{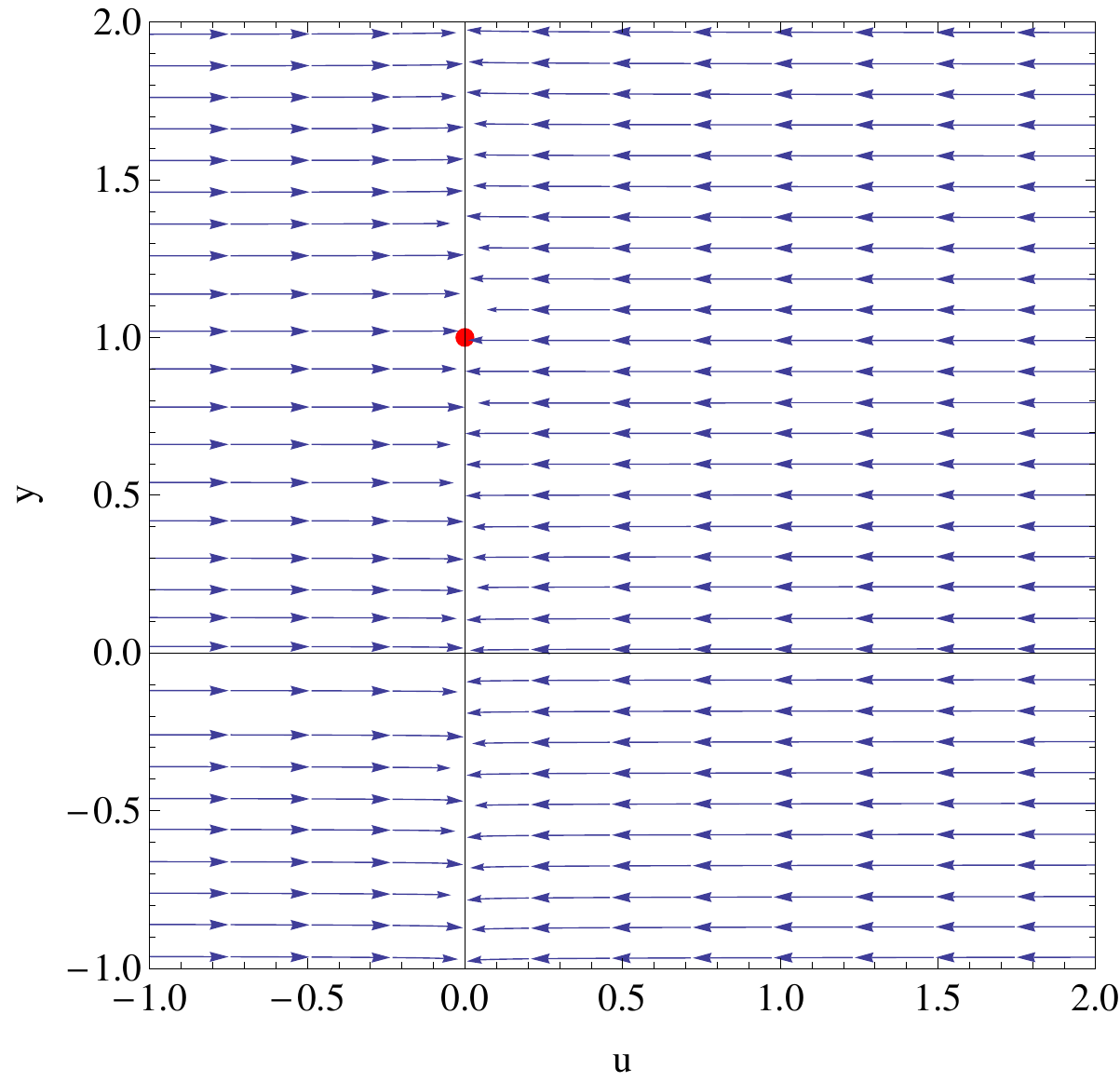}
 \caption{Phase space structure corresponding the critical points (1,0) (left panel) and (0,1) (right panel)} 
 \label{fig:ufig}
 \end{figure}
of this point would converge or diverge form the critical point depends on the initial condition. 
The geometrical structure of the phase space around this 
critical point is shown the in the first panel of figure.\ref{fig:ufig}, in which the saddle nature of the critical point 
is evident. It is to be noted that 
this critical point is corresponding to the solution at a prior decelerated epoch where dark matter is dominated. 
The saddle nature of the point is as expected such 
that the evolution will continue further towards a future epoch of the universe. \\

For the second critical point $(u_c,y_c)=(0,1)$
one of the eigen values is negative while the other is zero. It is to be noted that, the solution corresponding to 
this critical point is dark energy dominated at which the Hubble parameter is approaching a constant.
The phase plot of this given in the right panel of figure.\ref{fig:ufig} shows that, it is a non-isolated critical points. The plot shows that
the trajectories are attracted towards each other at the location of the critical points (0,1). Hence as far as 
the two dimensional plot is concerned the 
system is happens to be stable, which is the conclusion that is in compatible with the thermal evolution of the model. 

\section{Conclusions}

The cosmological model with dark matter and dynamical vacuum, $\Lambda=\Lambda_0+3\beta h^2$ has been analysed,
particularly in the context of a phenomenological interaction 
between the dark sectors. The model has been analysed before\cite{Bessada1} in the light of the cosmic star formation data and found a upper limit on the 
vacuum parameter as, $\beta<0.01$ In the present work we have analysed the cosmological evolution of the model using 
the supernovae data and estimated the 
dynamical vacuum density parameter as $\beta=-0.002$ and when known interaction between the dark sectors were characterized by a 
coupling constant, $b=0.001.$
It was found that the cosmological evolution of the model is almost similar in characteristics of the standard $\Lambda$CDM 
model.

The thermal evolution of the model has been studied and found that the generalized second law of entropy has been satisfied through out 
the evolution. The evolution of the second derivative of the entropy has also been analysed and found that it asymptotically approaches zero from the 
negative values, hence the condition for convexity, $\ddot S<0$ is satisfied. This shows that the model predicts a universe which resemble an ordinary 
macroscopic system in it's thermal behavior.

We have also performed a study on dynamical system behavior of the model, and found two critical points in the asymptotic conditions. The one 
corresponds to a solution in prior decelerated epoch of the universe and is a saddle point. Hence model will evolve from such a epoch to a future 
state. The later critical point is corresponding to an accelerating phase of the universe. The phase trajectories around this critical point as shown 
in figure \ref{fig:ufig}, shows that the trajectories are attracted towards each other form both sides of the critical point resembling a non-isolated 
point. However the approaching of the trajectories indicating a stable situation as far as the two dimensional behavior of phase-space is concerned.

 \end{document}